\documentstyle[aps,multicol,epsfig,prb]{revtex}
\def\etal{{\it et al.\/}}
\begin{document}
\title{Global Minima for Transition Metal Clusters Described by Sutton-Chen 
Potentials}
\author{Jonathan P.~K.~Doye}
\address{FOM Institute for Atomic and Molecular Physics, 
Kruislaan 407, 1098 SJ Amsterdam, The Netherlands}
\author{David J.~Wales}
\address{University Chemical Laboratory, Lensfield Road, Cambridge CB2 1EW, UK}
\date{\today}
\maketitle
\begin{abstract}
Using a Monte Carlo minimization approach we report
the global minima for metal clusters modelled by the Sutton-Chen family of
potentials with $N\le80$, where $N$ is the number of atoms. The resulting
structures are discussed in the light of both experimental and theoretical
data for clusters of the appropriate elements. 
\end{abstract}

\begin{multicols}{2}
\section{Introduction}
The structure is probably the most fundamental property of a cluster
and is important for understanding all aspects of chemical and physical 
behaviour.
Unfortunately, there is, as yet, no direct method for the structural
determination of free clusters in molecular beams.
Instead, one has to measure properties which depend upon geometry
and then try to infer the structure by comparing the results with the 
predictions of models.
This approach has been combined with techniques such as electron 
diffraction,\cite{Farges88} mass spectral abundances,\cite{Martin96}
chemical reactivity,\cite{Riley94} magnetism\cite{Bloomfield96}
and x-ray spectroscopy.\cite{Kakar97}

For transition metal clusters a wealth of structural information is 
now becoming available from increasingly sophisticated experiments. 
One of the most powerful techniques is the flow-reactor 
approach which probes the chemical reactivity of size-selected clusters. 
For example, this method has been applied to nickel 
clusters using nitrogen as the chemical probe to give detailed 
information for 
all sizes up to $N$=71.\cite{Parks94,Parks95a,Parks95b,Parks97}
From these data it has been possible to make structural assignments around 
$N$=13 and 55 (sizes of complete Mackay icosahedra\cite{Mackay}), but in
the size range $29\le N\le 48$ only one structural assignment has so far been 
made because of the large number of possible geometries to be considered 
and the presence of multiple isomers.\cite{Parks97}

It is, therefore, increasingly important that the theoretician
aids the task of experimental interpretation by producing
reliable structural models.
However, for transition metal clusters this is an extremely 
demanding task.
It is now becoming possible
to perform {\it ab initio\/} calculations for clusters at the 
larger end of the size range that we consider here 
($N\le 80$)\cite{Haberlan97,Jennison97}
but only for a few (usually high symmetry) geometries. It
is not feasible to perform the extensive
search of the potential energy surface required to 
find the most stable structure. 
Instead one has to use empirical potentials.
Even with these simplified descriptions of the interatomic interactions
it can be a very difficult task to find the global minimum for the 
size range considered here.
However, we are confident that in this work reliable estimates for the 
global minima have been found based upon the performance of the chosen algorithm
in previous studies\cite{WalesD97} and because of the large database of 
structures we have acquired through work on clusters bound
by the Morse potential.\cite{JD95c,JD97e} 

\begin{figure}
\begin{center}
\vglue -3mm
\epsfig{figure=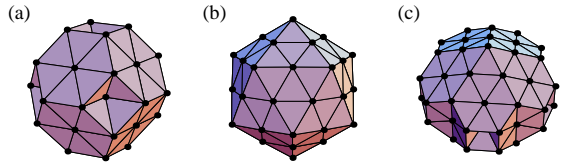,width=8.5cm}
\vglue -3mm
\begin{minipage}{8.5cm}
\caption{\label{fig:fid} (a) 38-atom truncated octahedron,
(b) 55-atom Mackay icosahedron, and (c) 75-atom Marks decahedron.
These clusters have optimal shapes for the three main types of 
ordered packing seen in this study: face-centred cubic (fcc), 
icosahedral and decahedral, respectively.
The latter two morphologies cannot be extended to the bulk
because they possess five-fold axes of symmetry.
All three structures are global minima for the SC 12--6 and 9--6 potentials,
and only the Mackay icosahedron is not the global minimum
for the 10--8 potential.}
\end{minipage}
\end{center}
\end{figure}

The three main morphologies that metal and simple atomic clusters 
adopt are icosahedra, decahedra and close-packing.
Particularly stable examples of each are given in Fig.\ \ref{fig:fid}.
The icosahedra and decahedra exhibit five-fold axes of symmetry which are
permitted due to the absence of translational periodicity.
The Mackay icosahedron\cite{Mackay} (Fig.\ \ref{fig:fid}b) can be decomposed
into twenty face-centred cubic (fcc) tetrahedra sharing a common vertex, 
and the decahedra are based upon pentagonal bipyramids 
made up of five fcc tetrahedra sharing a common edge.
The most stable decahedral form, the Marks decahedron\cite{Marks84} 
(Fig.\ \ref{fig:fid}c), 
involves further facetting to make the cluster more
spherical whilst keeping the proportion of the surface exhibiting
$\{100\}$ facets to a minimum. 
The most stable fcc cluster is the truncated octahedron.
All three morphologies are commonly seen in metal clusters supported
on metal surfaces.\cite{Marks94}

In this paper we seek to further our understanding of transition metal 
clusters and aid structural assignments from experimental data 
by performing global optimization for clusters interacting via the 
Sutton-Chen family of potentials for $N\le 80$.
In section \ref{methods} we describe the methods used and in section
\ref{results} we describe the structures of the global minima and 
compare them to experimental results and previous theoretical studies.

\section{Methods}
\label{methods}

\subsection{Potentials}

The Sutton-Chen (SC) potential has the form:\cite{Sutton90}
$$ E = \epsilon\sum_{i}\left[{1\over2}\sum_{j\not=i}
\left(a\over r_{ij}\right)^{n}-c\sqrt{\rho_{i}}\right],$$
$$\qquad {\rm where} \qquad
\rho_{i} = \sum_{j\not=i}\left(a\over r_{ij}\right)^{m}. $$
$c$ is a dimensionless parameter,
$\epsilon$ is a parameter with dimensions of energy, $a$ is the lattice constant,
and $m$ and $n$ are positive integers with $n>m$.
We use the $n$, $m$ and $c$ parameters given by Sutton and Chen\cite{Sutton90}\ 
for the metals Ag, Ni and Au; Rh has the same scaled parameters as Ag, 
Cu the same as Ni and 
Pt is the same as Au, so the corresponding results for
these metals can simply be obtained from their partners by rescaling.
For Ag and Rh $n = 12$, $m = 6$ and $c=144.41$,
for Ni and Cu $n = 9$, $m = 6$ and $c=39.432$,
for Au and Pt $n = 10$, $m = 8$ and $c=34.408$.
In the present calculations we employed reduced units with $\epsilon=1$ and $a=1$.
The tabulated results may therefore easily be rescaled for any of 
the above elements.
The appropriate energy is given simply by multiplying the reduced energy 
by $\epsilon$ whilst the coordinates need to be multiplied by 
$a$, i.e.~the lattice constant.
The Sutton-Chen potential provides a reasonable description of 
various bulk properties,\cite{Sutton90,Lynden91}\
with an approximate many-body representation of the delocalized metallic bonding.
However, it does not include any directional terms, which are likely 
to be important for transition metals with partially occupied $d$ shells.

\subsection{Global Optimization methods}

The main method we used to find the lowest minima is based upon Li and 
Scheraga's Monte Carlo minimization\cite{Li87a} or `basin-hopping' algorithm 
which we have recently explored for several systems.\cite{WalesD97,JD97e,JD97d}
This approach belongs to the family of `hypersurface deformation' 
methods\cite{StillW88} where the energy is transformed, 
usually to a smoother surface with fewer minima.
The lowest minimum of the new surface is then mapped back to the original
surface, but there is no guarantee that the global minima on the two surfaces
are related and often they are not.\cite{JD95c}\
In contrast, the transformation that we apply is guaranteed to preserve
the global minimum. The transformed energy $\tilde E$ is defined by:
$$
 \tilde E({\bf X}) = min\left\{ E({\bf X}) \right\},
$$
where ${\bf X}$ represents the vector of nuclear coordinates
and $min$ signifies that an energy minimization is performed starting
from ${\bf X}$.

The topography of the transformed surface is that of a multi-dimensional
staircase. Each step corresponds to the basin of attraction surrounding
a particular minimum (the set of geometries where
geometry optimization leads to that minimum).
The transformation has a significant effect on the dynamics.
Not only are transitions to a lower energy minimum barrierless, but they can
also occur at any point along the boundary between basins of attraction,
whereas on the untransformed surface transitions can occur only when the system
passes along the transition state valley.
As a result intrawell vibrational motion is removed and the system can hop
directly between minima at each step. The success of the basin-hopping method
for potential energy surfaces which exhibit multiple funnels has been explained 
elsewhere in terms of the thermodynamics of the transformed landscape.\cite{JD97d}
Similar methods have been used in studies of 
biomolecules.\cite{Baysal96,Derreumaux97a,Derreumaux97b}

To explore the $\tilde E$ surface we have used canonical Monte Carlo (MC)
sampling at temperatures of $T^*$=30, 5 and 10 for the 12--6, 10--8 and 9--6
potentials respectively, where the reduced temperature is $kT/\epsilon$.
To restrict the configuration space to bound clusters we 
reset the coordinates to those of the current minimum in the Markov chain at
each step.  
This objective can also be achieved by placing the cluster in a container.

In our recent application to LJ clusters the MC minimization approach
outperformed all other methods in the literature, finding all the known
lowest energy LJ clusters up to 110 atoms, including those with
non-icosahedral global minima.\cite{WalesD97}
All our published results are available in downloadable form from the
Cambridge Cluster Database.\cite{web}

In a recent study we found the global minima of 
clusters interacting with a Morse potential for all sizes 
up to 80 atoms as a function of the parameter in the Morse potential 
which determines the range of the interactions.\cite{JD95c,JD97e}
For this system there are at least 350 different global minima in this size range.
Most of the global minima have icosahedral, decahedral or closed-packed 
structures, and these were mainly found by considering the structures
that have the largest number of nearest-neighbour contacts 
for each morphology.\cite{JD95d}
To complement the basin-hopping calculations we reoptimized all the 
Morse global minima and low-lying structures for the SC potentials. 

\end{multicols}
\begin{figure}
\begin{center}
\epsfig{figure=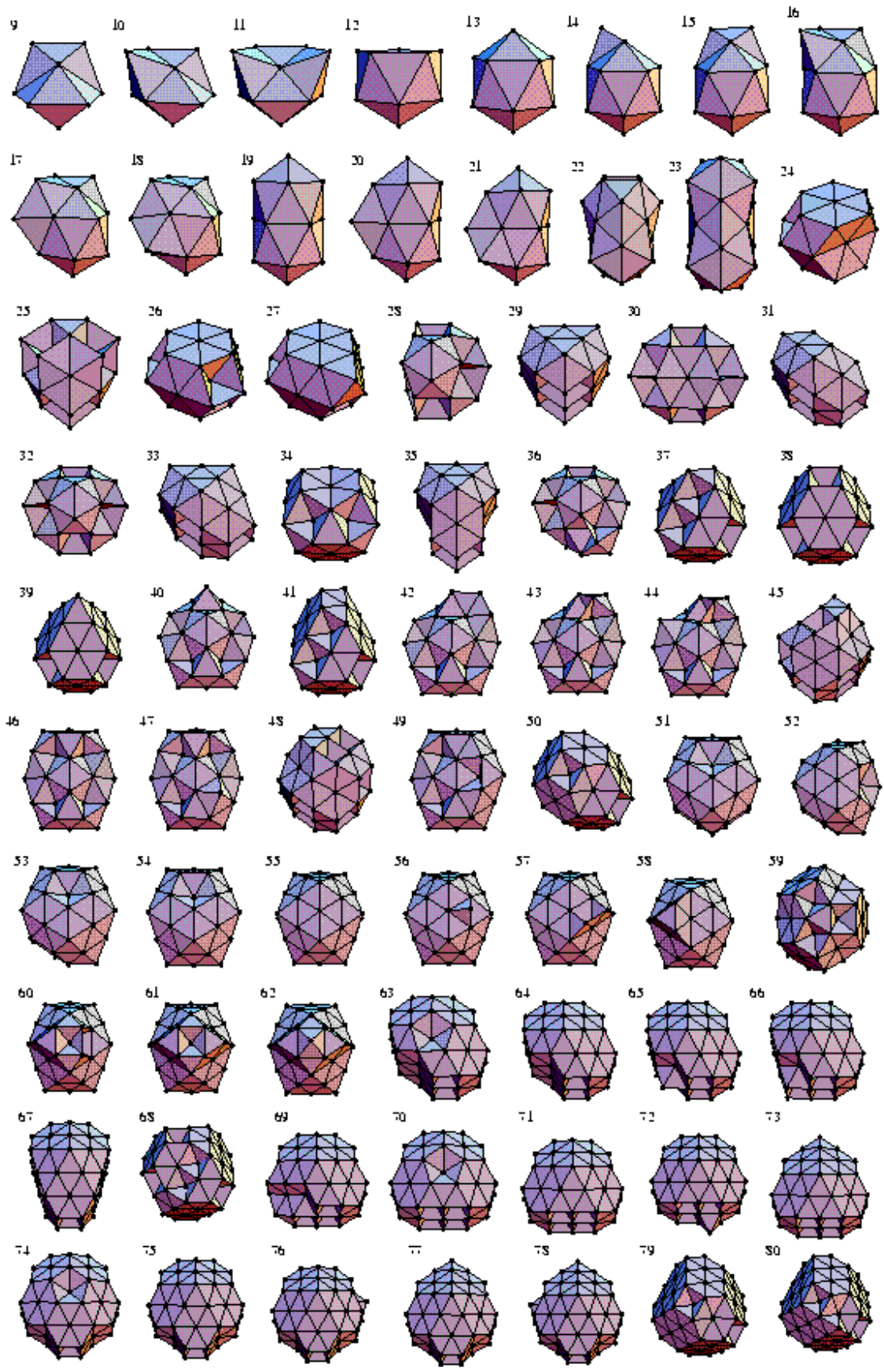,width=15cm}
\caption{\label{Ag:pics}Structures of the global minima for SC 12--6 clusters.}
\end{center}
\end{figure}
\begin{multicols}{2}

\section{Results}\label{results}
The energies and point groups of all the global minima are given 
in Table \ref{table}.
The basin-hopping algorithm found 95\% of the 12--6
global minima, 94\% of the 9--6 and 85\% of the 10--8. Five runs from different
random starting points consisting of 5000 MC steps
each were used for each cluster size.
These results generally confirm the utility and 
robustness of the basin-hopping approach.
Most of the failures occur when the global minimum has a close-packed
geometry which is only marginally lower in energy than an icosahedral or 
decahedral structure. 
In these cases the topography of the potential energy surface is likely to have 
multiple funnels, a scenario which often makes global optimization 
extremely difficult.\cite{JD97d,JD96c} We have not made any systematic attempt
to find the optimal temperatures or number of steps for the MC runs in the
present work.

It is interesting to note that 91\% of the 12--6 global minima are
also global minima for Morse clusters, 63\% of the 9--6 and 39\% of the 10--8.
The decreasing percentages reflect the relative propensity of these potentials
to give clusters with ordered structures of the icosahedral, decahedral or 
close-packed morphologies. These values also confirm our suggestion 
that the database of Morse global minima would be 
useful in providing candidate structures for studies 
with more sophisticated potentials.\cite{JD97e}
Furthermore, the results imply that some of the factors determining which isomer 
of a particular morphology is most stable are the same for a simple pair 
potential and for the more realistic many-body potentials used here.
Indeed virtually all the Sutton-Chen clusters that have an ordered icosahedral, 
decahedral or close-packed structure were found in the reoptimization of the 
larger Morse database (global minima and low-lying isomers), 
where the primary structural principle is the maximization of the 
number of nearest neighbours.

The results presented here match or better all previously published results for
Sutton-Chen clusters. 
We seem to have found the same structures as Nayak {\it et al.\/} 
who examined all 9--6 clusters with $N\le23$
(as no energies were given in that paper we cannot be sure of this).\cite{Nayak} 
Uppenbrink and Wales studied a selection of 12--6 and 10--8 clusters 
in terms of their thermodynamics and dynamics.\cite{Up93a}
They attempted to find the global minimum by performing 
regular minimizations from a molecular dynamics trajectory.
It is interesting to note that for the two larger sizes they considered 
($N$=40 and 55) the true global minimum was found in only one of the four cases.

\subsection{Silver and rhodium (SC 12--6) clusters}
\begin{figure}
\begin{center}
\epsfig{figure=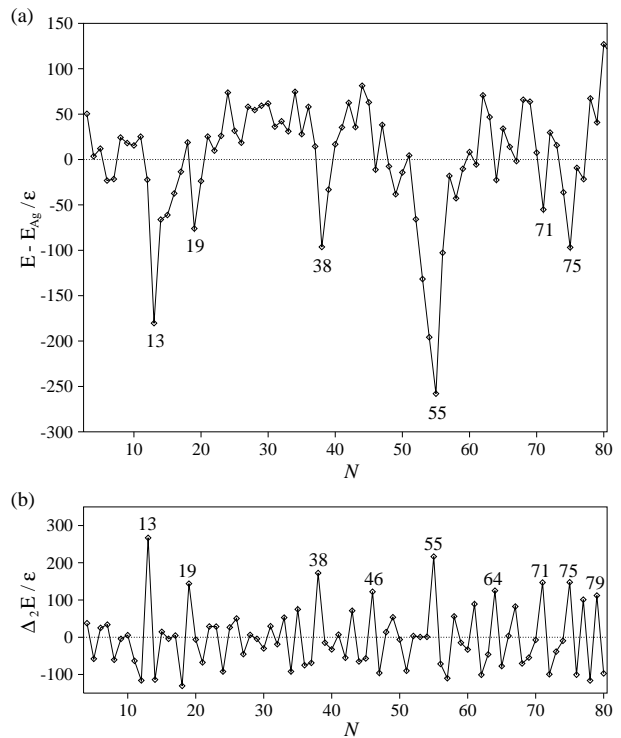,width=8.5cm}
\vglue 0.2cm
\begin{minipage}{8.5cm}
\caption{\label{Ag:energies}(a) Energies and (b) $\Delta_2 E$ for 
SC 12--6 clusters. 
In (a) the energy zero, 
$E_{Ag}=940.6021-994.8068N^{1/3}+1126.5506N^{2/3}-1201.3951N$,
where the coefficients have been chosen to give the best fit to the energies.
}
\end{minipage}
\end{center}
\end{figure}

The structures of the 12--6 global minima are illustrated in Fig.\ \ref{Ag:pics}
and the energies are plotted in
Fig.\ \ref{Ag:energies}a. A function of the form
$a+bN^{1/3}+cN^{2/3}+dN$, which has been fitted to the energy, has been 
subtracted to emphasize the size dependence.
In Fig.\ \ref{Ag:energies}b $\Delta_2 E(N)=E(N+1)+E(N-1)-2 E(N)$ 
is also illustrated; $\Delta_2 E$ measures the stability of 
a structure with respect to neighbouring sizes; peaks in $\Delta_2E$ 
have been found to correlate well with the magic numbers 
(sizes at which clusters are particularly abundant) 
observed in mass spectra.\cite{Clemenger}

From Fig.\ \ref{Ag:energies}a it can be seen that the most stable structures 
occur at sizes corresponding to complete Mackay icosahedra\cite{Mackay} 
($N$=13 and 55).
This result was expected; 
it has been estimated from comparisons of stable sequences of clusters 
that the Mackay icosahedra are lowest in energy for up to a 
few thousand atoms in this system.\cite{Up92} 
This result also agrees with an {\it ab initio\/} study of silver clusters where,
at the few sizes considered, icosahedra were
always lower than fcc structures.\cite{Jennison97}
The predominance of icosahedral morphologies might also be expected 
from the behaviour of the Lennard-Jones potential to which the current 
functional form bears some resemblance.
For Lennard-Jones clusters there are only 
four non-icosahedral global minima in the range $13\le N\le 80$.\cite{WalesD97}
However, for the SC 12--6 potential there are only 30 icosahedral 
global minima in this size range; at sizes between the complete
Mackay icosahedra other types of structure become lower in
energy. {\it Clearly the many-body part of the potential is
important in determining the most stable structure\/}.

Initially the growth sequence is similar to that of Lennard-Jones clusters.
From $N$=7 to 13 growth occurs by capping the 7-atom pentagonal bipyramid 
and leads to the 13-atom icosahedron. The one exception is the 8-atom cluster
for which a deltahedral dodecahedron is lower in energy than the 
capped pentagonal bipyramid. 
Growth on a Mackay icosahedron can occur in two ways. In the first, the
anti-Mackay overlayer, the atoms are added in sites which are hexagonal 
close-packed (hcp)
with respect to the twenty tetrahedra that make up the icosahedron; 
i.e.\ for the 13-atom icosahedron atoms are added to the centres of the 
faces and to the vertices. In the second, the Mackay overlayer, the atoms 
are added in sites which are fcc with respect to the underlying tetrahedra
which leads to the next Mackay icosahedron.
From $N$=14 to 21 growth occurs in the anti-Mackay layer, passing
through the stable 19-atom double icosahedron (Fig.\ \ref{Ag:energies}).
However, above this size there are many types of competing
structure and the global minimum changes frequently with size.
Not until $N$=51 are the structures again 
uniformly icosahedral leading to the complete Mackay icosahedron at 55 atoms, 
However, Mackay icosahedra with one and two faces missing
do give rise to the shallow minima in the energy plot 
(Fig.\ \ref{Ag:energies}a) at $N$=46 and 49. 

The two other morphologies exhibited by the global minima 
are decahedral and close-packed.
In Table 1 the close-packed clusters have been divided into
those that are hcp, those that are fcc and those that involve 
a mixture of stacking sequences and twin planes. The most
stable cluster in this intermediate size range is the 38-atom
truncated octahedron (Fig.\ \ref{Ag:energies}a).  
The stability of this structure has recently been recognized in both
theoretical\cite{JD95c,JD97e} and experimental\cite{Parks97} work.
Its shape is close to the ideal Wulff polyhedron, and
it is the only fcc structure that is the global minimum for the
Lennard-Jones potential in this size range.

For some of the decahedral clusters the (pseudo)-fivefold
axis is not always obvious; it is obscured by overlayers on the $\{111\}$
faces surrounding the axis at $N$=25, 30, 45 and 48 for this potential
(and at $N$=45, 48, 58 for the 9--6 potential).
Also, two global minima ($N$=22 and 23) do not belong to any of the 
ordered morphologies. The 23-atom structure is based on two distorted 
face-sharing icosahedra and has been found before for 
Morse clusters;\cite{JD97e} the 22-atom structure is similar.

Growth on the 55-atom Mackay icosahedron again begins in the 
anti-Mackay overlayer; this leads to the weak minima in the
energy plot of Fig.\ \ref{Ag:energies}a at $N$=58 and 61 which 
correspond to complete overlayers on one or two faces 
of the icosahedron.
However, decahedral structures soon become lower in energy. 
From $N$=63 a growth sequence begins
which leads to the 75-atom Marks decahedron. The latter
structure's stability is clear from Fig.\ \ref{Ag:energies}.

Unfortunately, it is difficult to make any critical
assessment of the performance and reliability of the 12--6
SC potential because there has been little theoretical or experimental
work on the structure of silver or rhodium clusters in the size 
range we consider here.

\subsection{Nickel and copper (SC 9--6) clusters}

\begin{figure}
\begin{center}
\epsfig{figure=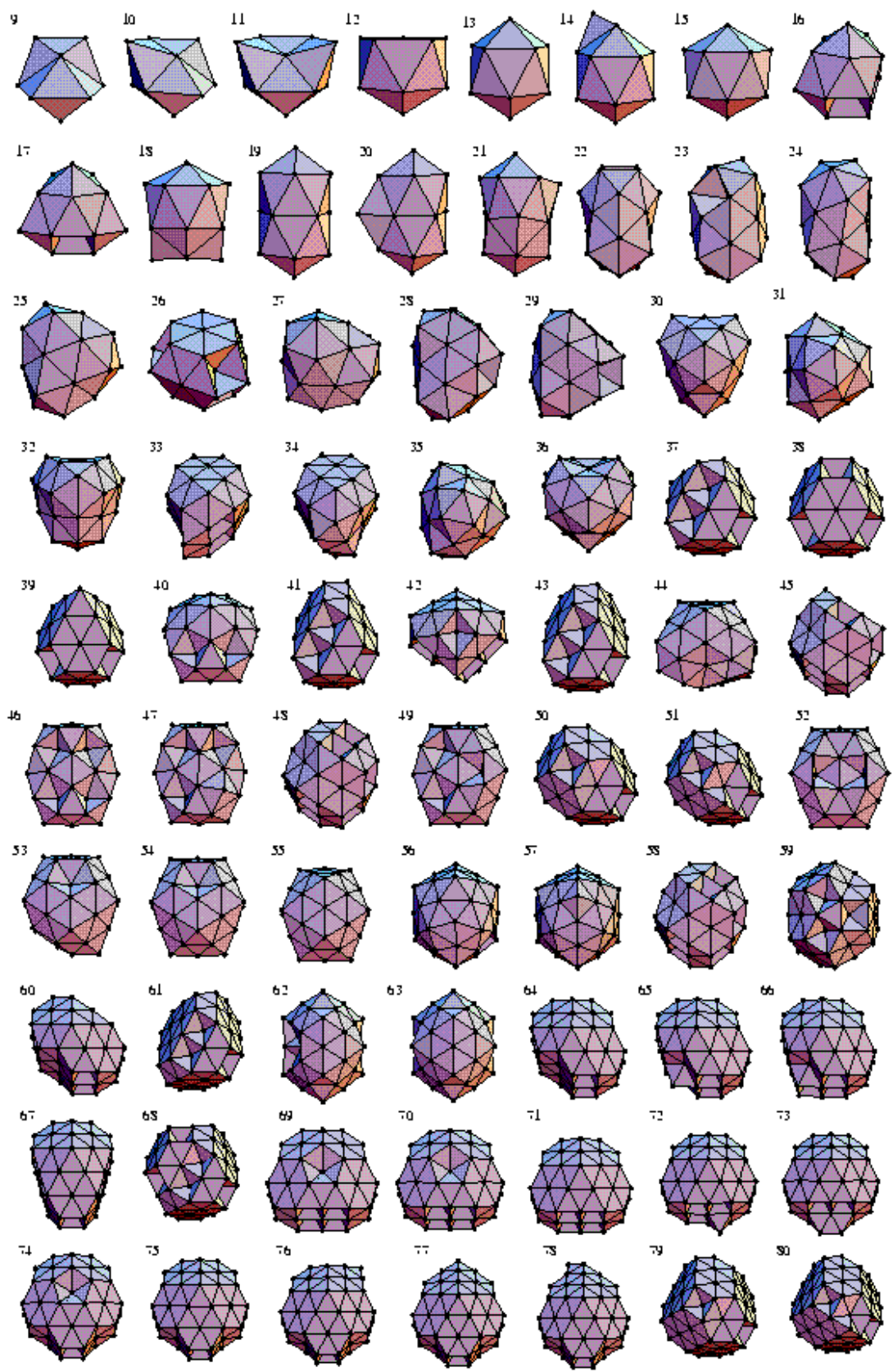,width=15cm}
\caption{\label{Ni:pics}Structures of the global minima for SC 9--6 clusters.}
\end{center}
\end{figure}

\begin{figure}
\begin{center}
\epsfig{figure=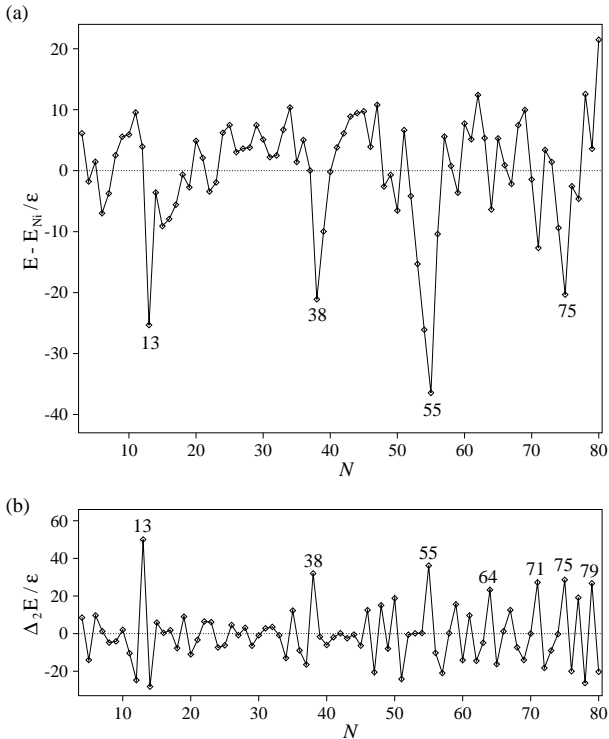,width=8.5cm}
\vglue 0.2cm
\begin{minipage}{8.5cm}
\caption{\label{Ni:energies} (a) Energies and (b) $\Delta_2 E$ for 
SC 9--6 clusters. 
In (a) the energy zero, 
$E_{Ni}=271.8994-292.8873N^{1/3}+260.6812N^{2/3}-292.9018N$,
where the coefficients have been chosen to give the best fit to the energies.
}
\end{minipage}
\end{center}
\end{figure}

The 9--6 global minima are illustrated in Fig.\ \ref{Ni:pics} and
the size-dependence of the energies and $\Delta_2 E$ are given
in Fig.\ \ref{Ni:energies}. The latter figure is very similar
to that found for the 12--6 clusters (Fig.\ \ref{Ag:energies}):
Mackay icosahedra are again most prominent, followed 
by the truncated octahedron and the Marks decahedron. 
However, the differences between the depths of the icosahedral ($N$=13, 55) and
non-icosahedral ($N$=38, 75) minima in the energy plot 
in Fig.\ \ref{Ni:energies}a have been reduced compared to 
Fig \ref{Ag:energies}a; 
this reflects a slight stabilization of the fcc and decahedral structures
with respect to the icosahedra. Moreover, there are no 
subsidiary minima in the energy plot due to icosahedral structures 
at $N$=19, 46 and 49. 
In total there are only 11 icosahedral minima in the size range 
$13\le N\le 80$.

Two additional close-packed structures at 
$N$=50 and 59 become more prominent in Fig.\ \ref{Ni:energies}.
The 50-atom cluster has $D_{3h}$ point group symmetry
and has a twin plane passing through the centre;
each half of the cluster on either side of this twin plane
has a structure which is a fragment of the 38-atom truncated 
octahedron. 
An analogous structure is the global minimum for $N$=79.
The 59-atom cluster is based on the 31-atom truncated tetrahedron 
with the four sides covered by seven-atom hexagonal overlayers.
Both are global minima for Morse clusters.\cite{JD97e}

Probably the biggest difference from the 12--6 clusters is the 
large number of global minima in the range $N=$16 to 44 which do not 
belong to any of the ordered morphologies. 
There are 25 such structures. 
The central atom of the 15-atom cluster is 14-coordinate. 
This structure is one of the Frank-Kasper coordination 
polyhedra,\cite{FrankK58,FrankK59} 
and is the global minimum for a long-ranged Morse potential.\cite{JD97e} 
A single negative disclination\cite{NelsonS} runs through this cluster.
The 16- and 17-atom structures are based on the 13-atom Ino decahedron
with three and four of the square faces capped, respectively. 
However, the structures
distort so that two caps approach to form a nearest-neighbour contact. 
They are related to the undistorted decahedra by a single 
diamond-square-diamond rearrangement.\cite{Lipscomb}
The structures for $N$=21--24 are similar to the 23-atom 12--6 global 
minimum which is made up of two distorted face-sharing icosahedra.

\end{multicols}
\begin{multicols}{2}

The 25-atom cluster is based on a $C_{3v}$ fcc structure, but with the 
triangular face twisted to remove $\{100\}$ facets. 
The 34-atom cluster resembles the decahedral 33-atom structure; however, the 
additional atom causes one part of the cluster to distort. 
The 40-atom global minimum is based on an icosahedral structure with a Mackay 
overlayer (as for 12--6). 
However, a low coordination number atom in that structure has been absorbed 
into the surface.
Similarly, the 56-atom and 57-atom clusters are based on the Mackay icosahedron. 
The first adatom adsorbs into the surface layer with an accompanying distortion,
rather than occupying a three-coordinate site on the surface.
The second adatom then occupies a four-coordinate site on the 
resulting distorted surface.
Similar avoidance of structures with atoms of low coordination number has been
seen in calculations for nickel by Wetzel and DePristo.\cite{Wetzel}
The 62- and 63-atom structures are based on the 52-atom global minimum, 
which is a Mackay icosahedron with an edge removed. 
Two triangular faces are added over this 
missing edge.

How can these relatively disordered structures become global minima, 
and why do they disappear at larger sizes? 
Probably because most of these geometries are more spherical, have a larger 
number of nearest-neighbour contacts and involve fewer low coordination number
atoms than the possible structures based on one of the ordered morphologies.
However, this advantage is counterbalanced by the introduction 
of considerable strain.
As the energetic penalty for the strain scales with the 
volume,\cite{JD95c,WalesD96}
it is only possible to accommodate it at smaller sizes.

For the elements described by this potential, there is a wealth of theoretical 
and experimental work to compare with our results. 
This fortunate situation is mainly due to the pioneering work of Riley 
and coworkers on nickel clusters.
First we compare our results to other theoretical
studies which use different descriptions of the interactions. 
Of these studies the most comprehensive is that of DePristo and
coworkers who used the corrected effective medium theory to look at all 
clusters up to $N$=55.\cite{Wetzel,Stave} 
They found that at $N$=13 and 55 the favoured geometries were icosahedral.
However, for many of the intermediate sizes, especially when the icosahedral 
structure at that size would involve a low coordination number surface atom, 
the global minima did not belong to any of the ordered morphologies. 
At some sizes these disordered structures appear to be the same as found 
in this study, e.g.~$N$=15, 18, and 25. 
DePristo {\it et al.\/} did not identify any of the global minima as 
close-packed or decahedral, although it seems that their 38-atom structure is in
fact a distorted truncated octahedron.\cite{Parks97}

Using a tight-binding model Lathiotakis \etal\ 
studied a selection of clusters with up to 55 atoms.\cite{Lathiotakis} 
Of the structures they considered, they found that Mackay icosahedra
were lowest in energy at $N$=13 and 55, but at some intermediate sizes
fcc structures were lower in energy than icosahedral clusters.
However, only a few structures 
were considered at each size and these may not include the global minima.

Montejano \etal\ used an embedded-atom potential to compare the energies
of icosahedral structures, in particular to locate the size at which the 
transition from an anti-Mackay to a Mackay overlayer occurred.\cite{Montejano}
However, these results are not relevant to many sizes since theory 
and experiment suggest that other non-icosahedral structures are lowest in
energy. 

Finally, Hu \etal\ attempted to model the interatomic interactions 
using a long-ranged Morse potential.\cite{Mei} 
As expected for a Morse potential 
with range parameter\cite{JD97e} $\rho_0$=3.54 they observed 
icosahedral structures with an anti-Mackay overlayer up to $N$=40. 
However, a Morse potential is probably not a good approximation to the real 
nickel potential;
therefore these results differ from experimental and other 
theoretical studies.

Icosahedral structure was first identified for nickel clusters 
by looking at the size dependence of the chemical reactivity with
various probe molecules.\cite{Winter91a,Parks91,Klots91}
Features were found at sizes corresponding to complete Mackay icosahedra, 
and also for icosahedral structures with stable surface overlayers.
Subsequently, icosahedral magic numbers have also been seen in mass spectra
for $50\le N\le 800$.\cite{Pellarin}

The most detailed information on the structure at particular sizes
has come from work using nitrogen probe 
molecules.\cite{Parks94,Parks95a,Parks95b,Parks97}
This technique has enabled structural assignments to be made for the majority
of clusters in the range $N\le 28$, $49\le N\le 71$ and at $N$=38.
However, in the intermediate size range, structural assignments based on 
these nitrogen experiments have not yet been published due to the 
difficulty in interpreting the experimental data.

In agreement with our work Parks \etal\ found that
the structures at $N$=13 and 55 are Mackay icosahedra, whilst
$N$=38 is a truncated octahedron.
In the range $13<N\le26$ the experiments with nitrogen indicate 
that most of the structures are formed by the
addition of an anti-Mackay overlayer to the 13-atom icosahedron,
from $N$=49 to 55 the Mackay overlayer is completed, and then up to $N$=71
an anti-Mackay overlayer grows on the 55-atom Mackay icosahedron.
The reactivity of nickel clusters with 
ammonia, water,\cite{Parks91} hydrogen\cite{Klots91} and deuterium\cite{Parks96a} 
also suggests that icosahedral structure persists above $N$=71.
These assignments differ from our 9--6 results in
that the Sutton-Chen potential seems to underestimate the stability of 
icosahedral structures.
However, non-icosahedral clusters do seem to occur 
in the range $29\le N\le 48$, in agreement with the SC 9--6 potential.
If disordered global minima are indeed observed in this size range 
for real nickel clusters this would help to explain the difficulties
experienced by Riley and coworkers in assigning structures.

To further facilitate comparisons with experiment we have estimated the number of 
binding sites for nitrogen using the rules formulated by Parks \etal\ 
These are: 
(1) N$_2$ binds directly to the nickel atoms; 
(2) a nickel atom with a coordination number of four or less binds two N$_2$ molecules;
(3) nickel atoms with a coordination number of five to eight will
readily bind one N$_2$ molecule; 
(4) nickel atoms with a coordination number of nine bind N$_2$ molecules 
weakly or not at all; and (5) nickel atoms with a coordination number of
ten or more do not bind N$_2$.
To clarify rule (4) it has been found that for smaller clusters N$_2$ can bind 
to a nine-coordinate atom, but for the larger clusters ($N>49$) no evidence has 
been found for this type of binding.
The number of binding sites and the coordination number analysis for the 9--6
global minima are given in Table \ref{table:N2}. 
For the clusters with an ordered morphology the differentiation between
nearest neighbours and next-nearest neighbours is clear.
However, this differentiation becomes ambiguous for the disordered 
structures and so the coordination number analysis, and sometimes
the number of N$_2$ binding sites, becomes
dependent on the nearest-neighbour criterion that is used.

The magnetic moments of size-selected nickel clusters have been measured
for all clusters with up to $N$=200.\cite{Bloomfield96} 
For small $N$ there is considerable variation of the magnetic moment with size. 
It was concluded that features at $N$=13 and 55 indicate icosahedral 
structure.\cite{Bloomfield96}
However, it is difficult to 
decipher the structural information that is contained in other 
features of the magnetic moment size-dependence.

Much less structural information is available for copper clusters.
Recent experiments suggest that icosahedra predominate up to $N\sim 2500$ 
atoms and above this fcc clusters are more prevalent.\cite{Reinhard97}
Although that study encompasses clusters far larger than we consider here, 
our results are not inconsistent with this finding.

\subsection{Gold and platinum (SC 10--8) clusters}

\begin{figure}
\begin{center}
\epsfig{figure=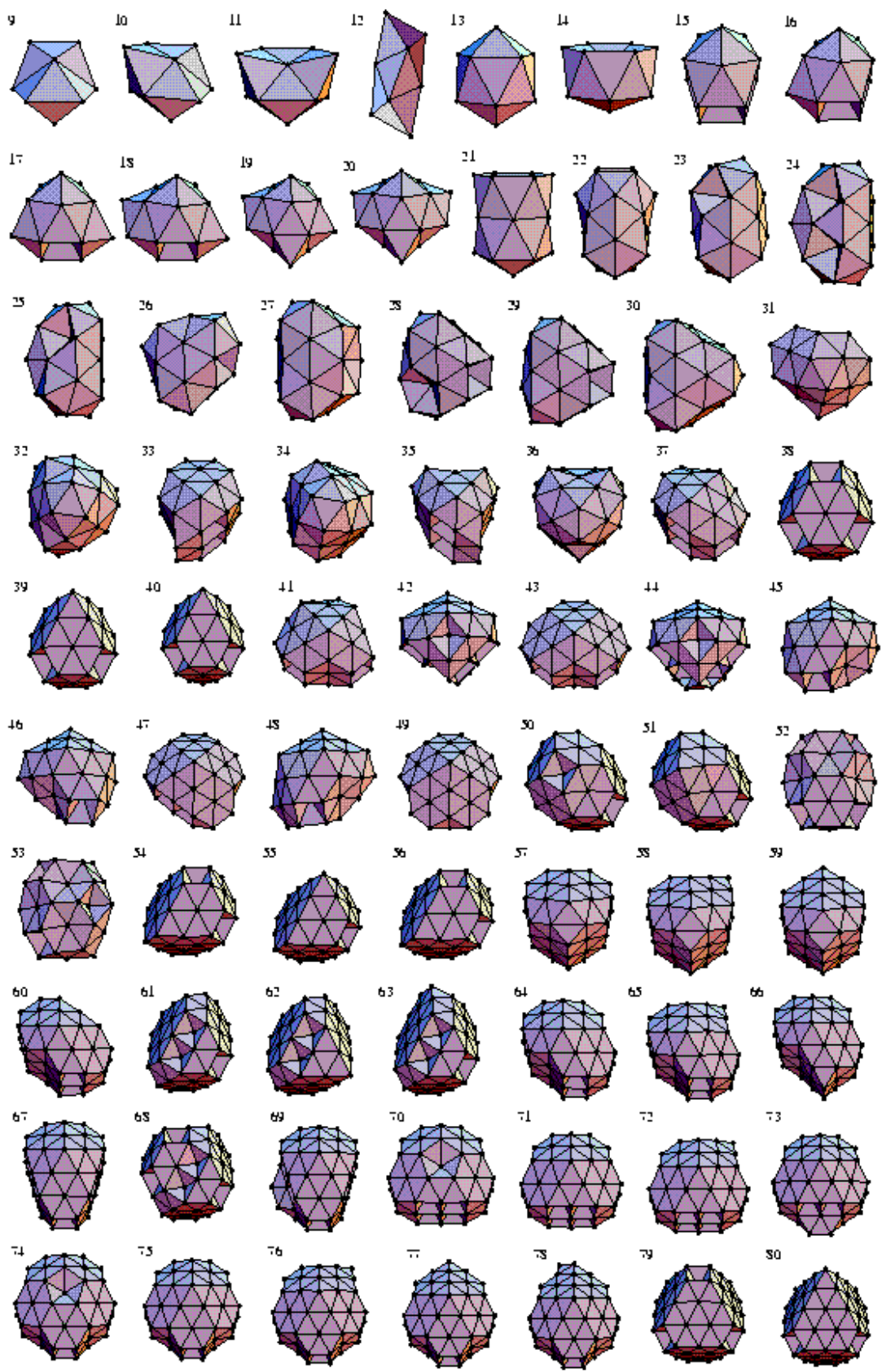,width=15cm}
\caption{\label{Au:pics}Structures of the global minima for SC 10--8 clusters.}
\end{center}
\end{figure}

\begin{figure}
\begin{center}
\epsfig{figure=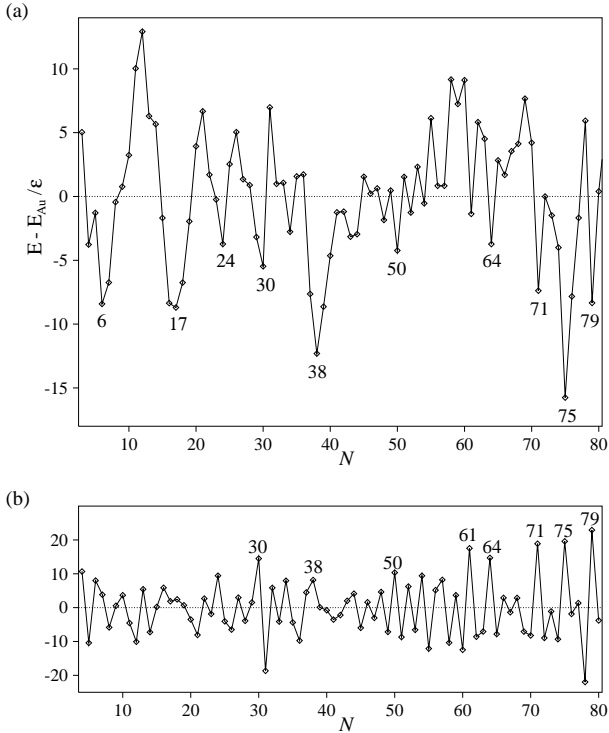,width=8.5cm}
\vglue 0.2cm
\begin{minipage}{8.5cm}
\caption{\label{Au:energies} (a) Energies and (b) $\Delta_2 E$ for 
SC 10--8 clusters. 
In (a) the energy zero, $E_{Au}=277.5753-276.0472N^{1/3}+192.0783N^{2/3}-305.9338N$,
where the coefficients have been chosen to give the best fit to the energies.
}
\end{minipage}
\end{center}
\end{figure}

The 10--8 global minima are illustrated in Fig.\ \ref{Au:pics} and
the size-dependence of the energies and $\Delta_2 E$ are given
in Fig.\ \ref{Au:energies}.
The latter figure shows a very different pattern from that seen for the 
12--6 or 9--6 clusters: there are no
signatures due to icosahedral clusters. Although the 13-atom
icosahedron is the global minimum, above this size there
are no global minima with ordered icosahedral structures.
Instead Fig.\ \ref{Au:energies}a is dominated by features
due to particularly stable close-packed and decahedral clusters.
The 38-atom fcc truncated octahedron and the 75-atom Marks
decahedron exhibit the deepest minima in Fig.\ \ref{Au:energies}a. 
There are also 
minima in the energy plot at $N$=64 and 71 due to incomplete Marks
decahedra, and at $N$=50, 61 and 79 due to close-packed clusters. 
The 50-atom global minima is the `twinned truncated octahedron' that
is the global minimum for the 12--6 an 9--6 potentials; 
the 61-atom structure is a fragment of the 79-atom twinned 
truncated octahedron that is the global minimum for the 12--6 and
9--6 potentials, and the 79-atom global minimum is a truncated octahedron.
That the 79-atom global minimum is fcc 
illustrates the 10--8 potential's greater preference 
for fcc structures rather than close-packed structures involving 
twin planes. For $N$=54--56 fcc structures are also lowest in energy and 
the 59-atom close-packed structure with $T_d$ symmetry is not the
global minimum in contrast to the 12--6 and 9--6 potentials.

As for the 9--6 potential, between $N$=13 and 55 there is a tendency to
form structures that do not fit neatly into one of the ordered morphologies.
In total there are 31 such structures compared to 25 for the 9--6 potential.
The 14- and 21-atom structures are based on the SC 9--6 15-atom global minimum.
The 14-atom cluster has one of the 7-coordinate vertices removed 
and the 21-atom structure is formed when 6 atoms are added to the faces
surrounding a vertex, thus extending the negative disclination line.\cite{JD97e}
The structures for $N$=15--20, like the $N$=15, 16 SC 9--6 global minima, 
are based on distorted decahedra. Similar distorted decahedra are also 
observed at $N$=35, 37 and 41.

\end{multicols}
\begin{multicols}{2}

The global minima for $N$=22--30 all structurally related.
The smaller clusters resemble the 23-atom $D_{3h}$ SC 12--6 global minimum and the
larger ones are related to a 30-atom structure which is made up of
three interpenetrating $D_{3h}$ units.
Interestingly, the 24- and 30-atom structures produce minima in the 
energy plot in Fig.\ \ref{Au:energies}a.

Many of the other disordered structures have, in part, the surface structure
of an incomplete Mackay icosahedron, but are distorted in various ways, 
e.g. $N$=31, 36, 42, 44, 46, 48, 52 and 53. The 53-atom structure
is based on the Mackay icosahedron with three adjacent vertices removed,
an atom added in the centre of this incomplete face and
accompanying distortions. The 52-atom structure is then formed simply by the removal
of one atom.

The main work to which our results can be compared is that
of Whetten and coworkers on gold clusters passivated by alkylthiolates. 
\cite{Whetten96,Alvarez97,Cleveland97a,Cleveland97b,Schaaff97}
In these experiments the clusters selectively form specific sizes, 
which were isolated by fractional crystallization. 
Each fraction has a narrow size distribution. 
The clusters presumably form these specific sizes because 
of their stability. Furthermore, it is thought that the 
passivating surface layer does not perturb the structure significantly. If this
is true then the observed structures reflect those of the free clusters.

Most of the cluster sizes formed were close to those
expected for truncated octahedra.\cite{Whetten96,Andres96}
For the smallest sizes detailed structural investigations
were made by comparing experimental x-ray diffraction patterns with those 
calculated from structural models. 
This comparison led to the identification of the fractions
with $N\sim 75$, 101 and 146 as Marks 
decahedra,\cite{Alvarez97,Cleveland97b} and those with $N\sim$ 225, 459 
as twinned truncated octahedra.\cite{Cleveland97a} 
Recently, a fraction with $N\sim 38$ has also been isolated; 
the diffraction patterns suggest that it is the fcc truncated 
octahedron.\cite{Schaaff97} 

Our results for SC 10--8 agree very well with these experimental data. 
The two clusters observed experimentally for $N\le 80$
are the two that we find to be the most stable, namely
the 38-atom truncated octahedron and the 75-atom Marks decahedron.

Cleveland {\it et al.\/} have performed a number of theoretical 
studies to interpret the experimental results on 
passivated gold clusters.\cite{Cleveland97a,Cleveland97b}
They used an embedded-atom potential and looked at several sequences 
of structures.
As in the present study, 
they found that truncated octahedra and Marks decahedra were most stable.
They also found that introducing a twin plane
into a truncated octahedron leads to an increase in energy,
e.g. for $N$=79 the fcc $O_h$ structure was slightly lower in energy
than the twinned $D_{3h}$ structure. 
This trend seems to disagree with the experimental results---at 
$N\sim$ 225, 459 the diffraction patterns seem to suggest that 
the structure has a twin plane.
Garzon and Posado-Amirillas also used the embedded-atom method to
look at a 55-atom cluster.\cite{Garzon96} 
They found that a disordered structure was lower in energy than the 
icosahedron.

We also note that at $N$=55 the fcc cuboctahedron
is not the global minimum; it actually lies $78.1\epsilon$ above the fcc global 
minimum. In connection with this comment, it is interesting 
that a recent reinvestigation of ligated 55-atom gold
clusters, which were originally thought to be cuboctahedral,\cite{Schmid81,Wallenberg85}
seems to disprove this structural assignment.\cite{Rapoport97}

The only calculations for platinum that are relevant to our results are by
Sachdev \etal\cite{Sachdev92,Sachdev93}\
For clusters with up to 60 atoms they found that the lowest energy structures 
were disordered. In particular at $N$=55 these structures were lower than
the icosahedron and cuboctahedron. These results have some overlap with ours
but it is not clear whether Sachdev \etal\  found the global minima for 
the larger sizes.

\section{Conclusions}

In this paper we have found the likely global minima for clusters with up to 80 atoms
whose interactions are described by the SC family of potentials.  
These potentials can be used to model silver, rhodium, nickel, copper,
gold and platinum clusters. The results are encouraging, thus confirming
the utility of these empirical potentials. 
For example, the most stable structures generally appear to agree with experiment.
For nickel clusters the 13-atom and 55-atom Mackay icosahedra, 
the 38-atom truncated octahedron and the 75-atom Marks decahedron are
particularly stable. The former three clusters have been unambiguously 
identified in experimental studies of chemical reactivity using nitrogen
as a probe molecule, and it would be interesting to see if extensions of these
experiments to $N$=75 could identify the Marks decahedron. For gold clusters
we find the 38-atom truncated octahedron and the 75-atom Marks decahedron
to be particularly stable; Whetten and coworkers have been able to isolate
both these clusters when passivated by 
alkylthiolates.\cite{Whetten96,Alvarez97,Cleveland97a,Cleveland97b,Schaaff97}

Our results should also be useful in aiding structural 
assignments from experimental data on size-selected clusters. 
However, it would be surprising if these empirical potentials
could accurately reproduce all the intricacies of a particular 
cluster growth sequence.
This is especially true when the energy differences between competing
structures are small. In such cases the character of the global minimum will 
be particularly sensitive to the accuracy of the potential.

When comparing the structures described here to experiment, aside from possible 
inaccuracies in the potential, it should also be remembered that 
the global minimum is only rigorously the free energy global minimum at 
absolute zero. 
At finite temperature entropic effects may play a role in determining the 
most stable structure. 
These entropic effects are most likely to be influential when
the energy gap between the global minimum and other low energy minima is small.
For example, it has been shown that for a 38-atom Lennard-Jones cluster the
structure changes from fcc to icosahedral as the temperature increases.\cite{JD97d} 
Similarly, for a 75-atom Morse cluster with a medium-ranged potential 
there is a transition from a Marks decahedron to icosahedral structures.\cite{JD95c} 
Both these transitions stem from the larger entropy of the icosahedra;
there are many icosahedral minima which have energies just above the
global minimum, whereas there is a large energy gap between the global minimum 
and the next lowest energy minimum with the same morphology.
However, probably a more common effect of temperature when energy differences between 
low energy structures are small will be the presence of multiple
isomers.

Finally, the results presented here further illustrate the power of the 
`basin-hopping' or Monte Carlo minimization global optimization algorithm.\cite{Li87a}
This method has enabled us to locate global minima with some confidence 
for systems with up to 240 degrees of freedom, some of which exhibit a multiple funnel 
potential energy surface topography.

\acknowledgements
D.J.W.\ is grateful to the Royal Society for financial support.
The work of the FOM Institute is part of the research program of
`Stichting Fundamenteel Onderzoek der Materie' (FOM)
and is supported by NWO (`Nederlandse Organisatie voor Wetenschappelijk Onderzoek').

\eject

\end{multicols}
\begin{table}
\caption{\label{table}Global minima for Sutton-Chen 12--6, 9--6 and 10--8 clusters
with $N\le 80$.
For each minimum the energy and point group are given and a structural
assignment made if possible.
The structural categories are:
icosahedral with an anti-Mackay (aM) or a Mackay overlayer (M); 
decahedral with $n$ atoms along the decahedral axis (d$n$); 
close-packed fcc (f), hcp (h), or 
involving a mixture of stacking sequences and twin planes (c); and 
clusters involving disclination lines (dis).
}       
\vglue 0.2cm
\begin{tabular}{cccccccccc}
 & \multicolumn{3}{c}{12--6} & \multicolumn{3}{c}{9--6} & \multicolumn{3}{c}{10--8} \\
\cline{2-4}\cline{5-7}\cline{8-10}
\noalign{\vspace{1pt}}
 $N$ & Energy/$\epsilon$ & PG & & Energy/$\epsilon$ & PG & & Energy/$\epsilon$ & PG & \\
\hline
 3 &  -1704.6905 & $D_{3h}$ &     &   -480.8560 & $D_{3h}$ &    &   -633.7771 & $D_{3h}$ & \\
 4 &  -2601.8447 & $T_d$    &     &   -709.5396 & $T_d$    &    &   -904.1153 & $T_d$    & \\
 5 &  -3461.3452 & $D_{3h}$ &     &   -929.7341 & $D_{3h}$ &    &  -1163.7670 & $D_{3h}$ & \\
 6 &  -4378.8875 & $O_h$    &     &  -1163.9640 & $O_h$    &    &  -1433.8252 & $O_h$    & \\
 7 &  -5271.2947 & $D_{5h}$ &     &  -1388.5116 & $D_{5h}$ &    &  -1695.8893 & $D_{5h}$ & \\
 8 &  -6129.7564 & $D_{2d}$ &     &  -1611.8509 & $D_{2d}$ &    &  -1954.1206 & $D_{2d}$ & \\
 9 &  -7048.7552 & $C_{2v}$ &     &  -1839.9790 & $C_{2v}$ &    &  -2218.1861 & $C_{2v}$ & \\
10 &  -7972.0971 & $C_{3v}$ &     &  -2072.2436 & $C_{3v}$ &    &  -2481.7019 & $C_{3v}$ & \\
11 &  -8889.9627 & $C_{2v}$ &     &  -2302.4939 & $C_{2v}$ &    &  -2741.5489 & $C_{2v}$ & \\
12 &  -9871.2458 & $C_{5v}$ &     &  -2543.1611 & $C_{5v}$ &    &  -3005.9274 & $C_2$    & \\
13 & -10968.5082 & $I_h$    &   M &  -2808.5765 & $I_h$    &  M &  -3280.3843 & $I_h$    &   M \\
14 & -11798.8479 & $C_{3v}$ &  aM &  -3023.9716 & $C_{3v}$ & aM &  -3549.4023 & $C_{6v}$ & dis \\
15 & -12742.9841 & $C_{2v}$ &  aM &  -3267.5300 & $D_{6d}$ & dis & -3825.6495 & $C_{2v}$ & \\
16 & -13672.6475 & $C_s$    &  aM &  -3505.2600 & $C_s$    &    &  -4101.6928 & $C_s$    & \\
17 & -14606.3231 & $C_2$    &  aM &  -3742.6166 & $C_{2v}$ &    &  -4371.8696 & $C_s$    & \\
18 & -15535.3810 & $C_s$    &  aM &  -3978.1268 & $C_{2v}$ &    &  -4640.1812 & $C_s$    & \\
19 & -16595.0561 & $D_{5h}$ &  aM &  -4221.3539 & $D_{5h}$ & aM &  -4906.0562 & $C_s$    & \\
20 & -17510.9209 & $C_{2v}$ &  aM &  -4455.5507 & $C_{2v}$ & aM &  -5171.2334 & $C_{2v}$ & \\
21 & -18433.0300 & $C_1$    &  aM &  -4700.7823 & $C_1$    &    &  -5439.9207 & $C_{6v}$ & dis \\
22 & -19422.7209 & $C_s$    &     &  -4949.3235 & $C_s$    &    &  -5716.6689 & $C_s$    & \\
23 & -20383.3977 & $D_{3h}$ &     &  -5191.4317 & $C_2$    &    &  -5990.7388 & $C_2$    & \\
24 & -21315.4208 & $C_{2v}$ &   h &  -5427.4229 & $C_2$    &    &  -6266.6622 & $C_s$    & \\
25 & -22339.6319 & $C_{3v}$ &  d3 &  -5670.7723 & $C_3$    &    &  -6533.1606 & $C_1$    & \\
26 & -23337.2211 & $D_{3h}$ &   h &  -5920.3488 & $D_{3h}$ &  h &  -6803.6959 & $C_s$    & \\
27 & -24284.3891 & $C_s$    &   h &  -6165.3671 & $C_s$    &    &  -7080.7248 & $C_s$    & \\
28 & -25276.9501 & $C_{3v}$ &   M &  -6411.2387 & $C_1$    &    &  -7354.7939 & $C_1$    & \\
29 & -26263.2779 & $C_{2v}$ &  d4 &  -6654.0358 & $C_2$    &    &  -7632.7382 & $C_2$    & \\
30 & -27253.8536 & $C_{2v}$ &  d3 &  -6903.2657 & $C_s$    &    &  -7909.1542 & $C_{3v}$ & \\
31 & -28274.4371 & $C_{2v}$ &  d4 &  -7153.4410 & $C_3$    &    &  -8171.0816 & $C_1$    & \\
32 & -29265.3320 & $C_{2v}$ &   M &  -7400.8234 & $D_{2d}$ &    &  -8451.6848 & $C_3$    & \\
33 & -30274.9603 & $C_{2v}$ &  d4 &  -7644.6441 & $C_s$    & d4 &  -8726.4506 & $C_s$    & d4 \\
34 & -31231.7697 & $C_{2v}$ &   c &  -7889.3674 & $C_2$    &    &  -9005.3502 & $C_3$    & \\
35 & -32280.3945 & $C_{2v}$ &  d4 &  -8147.0475 & $D_3$    &    &  -9276.2927 & $C_s$    & \\
36 & -33253.9352 & $C_s$    &   M &  -8392.4962 & $C_{2v}$ &    &  -9551.6256 & $C_{2v}$ & \\
37 & -34302.6067 & $C_{3v}$ &   c &  -8646.8835 & $C_{3v}$ &  c &  -9836.6867 & $C_{2v}$ & \\
38 & -35419.9804 & $O_h$    &   f &  -8917.7056 & $O_h$    &  f & -10117.2454 & $O_h$    &  f \\
39 & -36364.8587 & $C_{4v}$ &   f &  -9156.5715 & $C_{4v}$ &  f & -10389.6477 & $C_{4v}$ &  f \\
40 & -37324.3708 & $C_s$    &   M &  -9397.0850 & $C_s$    &    & -10661.9303 & $D_{4h}$ &  f \\
41 & -38316.5698 & $C_s$    &   c &  -9643.6606 & $C_s$    &  c & -10934.9766 & $C_s$    & \\
42 & -39301.6696 & $C_s$    &   M &  -9892.1913 & $C_s$    &    & -11211.5400 & $C_s$    & \\
43 & -40341.8543 & $C_s$    &   M & -10140.5484 & $C_s$    &    & -11490.3063 & $C_{2v}$ & d4 \\
44 & -41310.9157 & $C_1$    &   M & -10391.3783 & $C_2$    &    & -11767.0685 & $C_s$    & \\
45 & -42345.0912 & $C_s$    &  d4 & -10642.7040 & $C_s$    & d4 & -12039.6977 & $C_1$    & \\
46 & -43436.2827 & $C_{2v}$ &   M & -10900.4123 & $C_{2v}$ &  M & -12318.3028 & $C_3$    & \\
47 & -44405.1884 & $C_1$    &   M & -11145.6538 & $C_1$    &  M & -12595.3291 & $C_{2v}$ & d4 \\
48 & -45470.1069 & $C_{2v}$ &  d4 & -11411.4049 & $C_{2v}$ & d4 & -12875.3949 & $C_1$    & \\
49 & -46521.2131 & $C_{3v}$ &   M & -11662.0840 & $C_{3v}$ &  M & -13150.8235 & $D_{5h}$ & d4 \\
50 & -47518.6719 & $D_{3h}$ &   c & -11920.7434 & $D_{3h}$ &  c & -13433.4182 & $D_{3h}$ &  c \\
51 & -48522.4267 & $C_{2v}$ &   M & -12160.5960 & $C_s$    &  c & -13705.6730 & $C_s$    &  c \\
52 & -49616.1377 & $C_{3v}$ &   M & -12424.6351 & $C_{2v}$ &  M & -13986.6288 & $C_{2v}$ & \\
53 & -50706.4665 & $C_{2v}$ &   M & -12689.2438 & $C_{2v}$ &  M & -14261.3449 & $C_{3v}$ & \\
54 & -51796.0777 & $C_{5v}$ &   M & -12953.6990 & $C_{5v}$ &  M & -14542.6311 & $C_{2v}$ &  f \\
55 & -52884.6806 & $I_h$    &   M & -13217.8963 & $I_h$    &  i & -14814.5225 & $C_1$    &  f \\
\end{tabular}
\end{table}
\eject
\addtocounter{table}{-1}
\begin{table}
\caption{continued.}
\vglue 0.2cm
\begin{tabular}{cccccccccc}
 & \multicolumn{3}{c}{12--6} & \multicolumn{3}{c}{9--6} & \multicolumn{3}{c}{10--8} \\
\cline{2-4}\cline{5-7}\cline{8-10}
\noalign{\vspace{1pt}}
 $N$ & Energy/$\epsilon$ & PG & & Energy/$\epsilon$ & PG & & Energy/$\epsilon$ & PG & \\
\hline
56 & -53756.6516 & $C_{3v}$ &   M & -13445.8961 & $C_s$    &    & -15098.5064 & $D_{2h}$ &  f \\
57 & -54700.1733 & $C_s$    &  aM & -13684.1489 & $C_s$    &    & -15377.3233 & $C_{2v}$ & d5 \\
58 & -55753.8515 & $C_{3v}$ &  aM & -13943.4015 & $D_{3h}$ & d4 & -15647.9269 & $C_s$    & d5 \\
59 & -56751.4572 & $T_d$    &   c & -14202.4032 & $T_d$    &  c & -15928.8952 & $C_{2v}$ & d5 \\
60 & -57763.6760 & $C_s$    &  aM & -14445.8412 & $C_{2v}$ & d5 & -16206.2020 & $C_{2v}$ & d5 \\
61 & -58809.0448 & $C_{2v}$ &  aM & -14703.3954 & $C_{3v}$ &  c & -16495.9638 & $C_{3v}$ &  c \\
62 & -59765.2180 & $C_{2v}$ &  aM & -14951.2644 & $C_s$    &    & -16768.1774 & $C_s$    &  c \\
63 & -60822.3826 & $C_s$    &  d5 & -15213.6371 & $C_{2v}$ &    & -17048.9913 & $C_s$    &  c \\
64 & -61925.6244 & $C_{2v}$ &  d5 & -15480.8530 & $C_{2v}$ & d5 & -17336.8555 & $C_{2v}$ & d5 \\
65 & -62903.7387 & $C_{2v}$ &  d5 & -15724.8185 & $C_{2v}$ & d5 & -17610.0214 & $C_{2v}$ & d5 \\
66 & -63959.3105 & $C_s$    &  d5 & -15985.0232 & $C_s$    & d5 & -17890.9860 & $C_s$    & d5 \\
67 & -65011.2767 & $C_{2v}$ &  d5 & -16244.0284 & $C_{2v}$ & d5 & -18169.0668 & $C_{2v}$ & d5 \\
68 & -65980.5983 & $C_{3v}$ &   c & -16490.5019 & $C_{3v}$ &  c & -18448.5177 & $C_{3v}$ &  c \\
69 & -67020.4042 & $C_1$    &  d5 & -16744.2741 & $C_{2v}$ & d5 & -18725.1157 & $C_1$    & d5 \\
70 & -68114.9462 & $C_s$    &  d5 & -17012.0775 & $C_s$    & d5 & -19008.8021 & $C_s$    & d5 \\
71 & -69216.6518 & $C_{2v}$ &  d5 & -17279.8708 & $C_{2v}$ & d5 & -19300.7255 & $C_{2v}$ & d5 \\
72 & -70171.4663 & $C_1$    &  d5 & -17520.4788 & $C_1$    & d5 & -19573.7638 & $C_s$    & d5 \\
73 & -71225.8547 & $C_{2v}$ &  d5 & -17779.2971 & $C_s$    & d5 & -19855.7668 & $C_s$    & d5 \\
74 & -72318.7243 & $C_{5v}$ &  d5 & -18047.0929 & $C_{5v}$ & d5 & -20138.8921 & $C_{5v}$ & d5 \\
75 & -73421.0521 & $D_{5h}$ &  d5 & -18315.1577 & $D_{5h}$ & d5 & -20431.3452 & $D_{5h}$ & d5 \\
76 & -74375.6975 & $C_s$    &  d5 & -18554.6196 & $C_{2v}$ & d5 & -20704.2050 & $C_{2v}$ & d5 \\
77 & -75430.9852 & $C_{2v}$ &  d5 & -18814.0659 & $C_{2v}$ & d5 & -20978.9269 & $C_{2v}$ & d5 \\
78 & -76385.4318 & $C_1$    &  d5 & -19054.3977 & $C_s$    & d5 & -21252.2801 & $C_s$    & d5 \\
79 & -77456.0255 & $D_{3h}$ &   c & -19321.0094 & $D_{3h}$ &  c & -21547.5979 & $O_h$    &  f \\
80 & -78414.6271 & $C_s$    &   c & -19560.8966 & $C_s$    &  c & -21819.9952 & $C_{4v}$ &  f \\
\end{tabular}
\end{table}
\begin{table}
\caption{\label{table:N2}Coordination number analysis and estimated
number of N$_2$ binding sites for SC 9--6 clusters.
The number of binding sites has been calculated using the rules given in the text. 
The values in brackets are appropriate if nine-coordinate binding sites are included. 
For 38 atoms and above there is no evidence of this type of binding and 
the alternative values have been omitted.
The nearest-neighbour criterion used is 0.8$a$.}
\vglue 0.2cm
\begin{tabular}{cccccccccc}
  & N$_2$ binding  &  \multicolumn{8}{c}{coordination number} \\
\cline{3-10}
\noalign{\vspace{1pt}}
$N$  & sites & 3 &  4 &  5 &  6 &  7 &  8 &  9 & $\ge 10$ \\
\hline
4  &   8 & 4 &    &    &    &    &    &    &    \\
5  &  10 & 2 &  3 &    &    &    &    &    &    \\
6  &  12 &   &  6 &    &    &    &    &    &    \\
7  &  12 &   &  5 &  2 &    &    &    &    &    \\
8  &  12 &   &  4 &  4 &    &    &    &    &    \\
9  &  13 &   &  4 &  2 &  2 &    &  1 &    &    \\
10 &  12(13) &   &  3 &  3 &  3 &    &    &  1 &    \\
11 &  12 &   &  2 &  4 &  4 &    &    &    &  1 \\
12 &  11 &   &    &  5 &  6 &    &    &    &  1 \\
13 &  12 &   &    &    & 12 &    &    &    &  1 \\
14 &  14 &   &    &    &  9 &  3 &    &    &  1 \\
15 &  14 &   &    &    & 12 &  2 &    &    &  1 \\
16 &  16 &   &  1 &  2 & 12 &    &    &    &  1 \\
17 &  18 &   &  2 &    & 10 &  4 &    &    &  1 \\
18 &  19 &   &  2 &    &  8 &  7 &    &    &  1 \\
19 &  17 &   &    &    & 12 &    &  5 &    &  2 \\
20 &  19 &   &  1 &    & 10 &  2 &  5 &    &  2 \\
21 &  19 &   &    &  1 & 12 &  1 &  5 &    &  2 \\
22 &  20 &   &    &  3 & 11 &  4 &  2 &    &  2 \\
23 &  21 &   &    &    & 14 &  6 &  1 &    &  2 \\
24 &  22 &   &    &  2 & 10 &  6 &  4 &    &  2 \\
25 &  21(22) &   &    &    & 12 &  3 &  6 &  1 &  3 \\
26 &  21(24) &   &    &  6 & 12 &    &  3 &  2 &  3 \\
27 &  24 &   &    &    & 13 &  4 &  7 &    &  3 \\
28 &  25 &   &    &    & 13 &  7 &  5 &    &  3 \\
29 &  26 &   &    &  2 & 12 &  6 &  6 &    &  3 \\
30 &  26 &   &    &  1 & 13 &  4 &  8 &    &  4 \\
31 &  24(27) &   &    &  3 & 12 &  3 &  6 &  3 &  4 \\
32 &  24(28) &   &    &    & 12 &    &    &  4 &  4 \\
33 &  26(26) &   &    &  4 & 14 &  2 &  6 &  2 &  5 \\
34 &  26(30) &   &    &  2 & 14 &  6 &  4 &  4 &  4 \\
35 &  30 &   &    &    & 12 &  6 &  8 &    &  5 \\
36 &  30 &   &    &    & 16 &  6 &  8 &    &  6 \\
37 &  27(31) &   &    &  3 & 18 &  3 &  3 &  4 &  6 \\
38 &  24 &   &    &    & 24 &    &    &  8 &  6 \\
39 &  26 &   &  1 &    & 20 &  4 &    &  8 &  6 \\
40 &  33 &   &    &    & 14 &  6 & 13 &    &  7 \\
41 &  29 &   &    &  3 & 18 &  4 &  4 &  5 &  7 \\
42 &  35 &   &    &  2 & 18 &  1 & 14 &    &  7 \\
43 &  29 &   &    &  3 & 18 &  4 &  4 &  7 &  7 \\
44 &  35 &   &    &    & 14 & 10 & 11 &    &  9 \\
45 &  32 &   &    &  2 & 19 &  5 &  6 &  3 & 10 \\
46 &  35 &   &    &    & 16 &  2 & 17 &  2 &  9 \\
47 &  36 &   &    &  1 & 16 &  2 & 17 &  2 &  9 \\
48 &  35 &   &    &    & 22 &  2 & 11 &  2 & 11 \\
49 &  36 &   &    &    & 15 &  3 & 18 &  3 & 10 \\
50 &  30 &   &    &    & 24 &  6 &    &  8 & 12 \\
51 &  32 &   &  1 &    & 22 &  7 &  1 &  8 & 12 \\
52 &  39 &   &    &    & 14 &  4 & 21 &    & 13 \\
53 &  40 &   &    &    & 10 & 10 & 20 &    & 13 \\
54 &  41 &   &    &    & 11 &  5 & 25 &    & 13 \\
55 &  42 &   &    &    & 12 &    & 30 &    & 13 \\
\end{tabular}
\end{table}
\addtocounter{table}{-1}
\begin{table}
\caption{continued.}
\vglue 0.2cm
\begin{tabular}{cccccccccc}
  & N$_2$ binding  &  \multicolumn{8}{c}{coordination number} \\
\cline{3-10}
\noalign{\vspace{1pt}}
$N$  & sites & 3 &  4 &  5 &  6 &  7 &  8 &  9 & $\ge 10$ \\
\hline
56 &  43 &   &    &    & 10 & 11 & 22 &    & 13 \\
57 &  45 &   &  1 &    &  9 &  9 & 25 &    & 13 \\
58 &  39 &   &    &    & 24 &  6 &  9 &    & 19 \\
59 &  36 &   &    &    & 24 & 12 &    &  4 & 19 \\
60 &  41 &   &    &    & 22 &    & 13 &  4 & 15 \\
61 &  36 &   &    &    & 24 &  6 &  6 & 12 & 13 \\
62 &  46 &   &    &    & 15 &  3 & 28 &    & 16 \\
63 &  46 &   &    &    & 16 &    & 32 &    & 17 \\
64 &  40 &   &    &    & 22 &  8 & 10 &  6 & 18 \\
65 &  42 &   &  1 &    & 22 &  6 & 12 &  6 & 18 \\
66 &  42 &   &    &  2 & 20 &  7 & 13 &  6 & 18 \\
67 &  44 &   &    &    & 22 &  6 & 16 &  6 & 17 \\
68 &  39 &   &    &    & 24 &  9 &  6 & 13 & 16 \\
69 &  41 &   &    &    & 20 & 18 &  3 &  8 & 18 \\
70 &  42 &   &    &    & 21 & 13 &  8 &  8 & 19 \\
71 &  43 &   &    &    & 22 &  8 & 13 &  8 & 20 \\
72 &  43 &   &  1 &    & 21 &  8 & 12 & 10 & 20 \\
73 &  41 &   &    &  2 & 20 &  9 & 10 & 11 & 21 \\
74 &  41 &   &    &    & 21 & 15 &  5 & 10 & 23 \\
75 &  42 &   &    &    & 22 & 10 & 10 & 10 & 23 \\
76 &  44 &   &  1 &    & 18 & 14 & 10 & 10 & 23 \\
77 &  44 &   &    &  2 & 18 & 12 & 12 & 10 & 23 \\
78 &  46 &   &  1 &    & 20 & 10 & 14 & 10 & 23 \\
79 &  39 &   &    &    & 24 & 12 &  3 & 18 & 22 \\
80 &  41 &   &  1 &    & 22 & 13 &  4 & 18 & 22 \\
\end{tabular}
\end{table}

\end{document}